\documentclass[manuscript]{aastex62}
\usepackage{color, soul}
\usepackage{graphicx}
\usepackage{subfigure}
\usepackage{rotating}
\usepackage{amsmath}

\shortauthors{Jin et al.}

\begin{document}

\title{{\it Chandra} Detection of Intra-cluster X-ray Sources in Fornax}

\correspondingauthor{Zhiyuan Li}
\email{xiangyu.jin@mail.mcgill.ca, lizy@nju.edu.cn}

\author{Xiangyu Jin}
\affiliation{School of Astronomy and Space Science, Nanjing University, Nanjing 210023, China}
\affiliation{School of Physics, Nanjing University, Nanjing 210023, China}
\affiliation{Department of Physics and McGill Space Institute, McGill University, 3600 University St., Montreal QC, H3A 2T8, Canada}

\author{Meicun Hou}
\affiliation{School of Astronomy and Space Science, Nanjing University, Nanjing 210023, China}
\affiliation{Key Laboratory of Modern Astronomy and Astrophysics (Nanjing University), Ministry of Education, Nanjing 210023, China}

\author{Zhenlin Zhu}
\affiliation{School of Astronomy and Space Science, Nanjing University, Nanjing 210023, China}
\affiliation{Key Laboratory of Modern Astronomy and Astrophysics (Nanjing University), Ministry of Education, Nanjing 210023, China}

\author{Zhiyuan Li}
\affiliation{School of Astronomy and Space Science, Nanjing University, Nanjing 210023, China}
\affiliation{Key Laboratory of Modern Astronomy and Astrophysics (Nanjing University), Ministry of Education, Nanjing 210023, China}

\begin{abstract}
Based on archival {\it Chandra} observations with a total exposure of 1.3 Ms, we study X-ray point sources in the Fornax cluster of galaxies, with the primary aim of searching for intra-cluster X-ray source populations. We detect 1177 point sources out to a projected radius of $\sim$30 arcmin ($\sim$180 kpc) from the cluster center and down to a limiting 0.5--8 keV luminosity of $\sim3\times10^{37}{\rm~erg~s^{-1}}$. We construct source surface density profile, after excluding sources associated with foreground stars, known globular clusters, ultra-compact dwarfs and galactic nuclei.  From this profile we statistically identify $\sim$183 excess sources that are not associated with the bulk stellar content of the individual member galaxies of Fornax, nor with the cosmic X-ray background. Taking into account Poisson error and cosmic variance, the cumulative significance of this excess is at $\gtrsim 2\,\sigma$ level (with a maximum of 3.6\,$\sigma$) outside three effective radii of the central giant elliptical, NGC\,1399. The luminosity function of the excess sources is found to be significantly steeper than that of the GC-hosting sources (presumably low-mass X-ray binaries [LMXBs]), disfavoring the possibility that unidentified GCs are primarily responsible for the excess. We show that a large fraction of the excess can be related to the extended stellar halo of NGC\,1399 and/or the diffuse intra-cluster light, thus providing strong evidence for the presence of intra-cluster X-ray sources in Fornax, the second unambiguous case for a galaxy cluster after Virgo.  Other possible origins of the excess, including supernova-kicked LMXBs and stripped nucleated dwarf galaxies are discussed.

\end{abstract}


\section{Introduction} \label{sec:intro}
X-ray-emitting, close binary systems involving an accreting black hole (BH) or neutron star (NS), are among the first objects discovered in the X-ray sky and now recognized to be ubiquitous in the local Universe, thanks in particular to {\it Chandra} observations \citep{Weisskopf2002} of excellent angular resolution and sensitivity.
As such, X-ray binaries can serve as a useful tool to study their parent stellar populations (see review by \citealp{Fabbiano2006}).
It is now a consensus that high-mass X-ray binaries (HMXBs) are prevalent in star-forming, typically late-type galaxies, whereas low-mass X-ray binaries (LMXBs) are the predominant population of detected X-ray sources (typically with luminosities $L_{\rm X} \gtrsim 10^{36}{\rm~erg~s^{-1}}$) in early-type galaxies (ETGs). 
This has motivated observational work to calibrate a quasi-linear relation between the total number (or total X-ray luminosity) of LMXBs and host galaxy's stellar mass \citep[e.g.,][]{Gilfanov2004MNRAS,KimFabbiano2004,Zhang2012}. 

In recent years, there has been growing evidence that X-ray sources also exist beyond the bulk stellar content of ETGs. 
Based on {\it Chandra} observations, \citet{Li2010ApJ} identified a significant ``excess" of X-ray sources at projected distances of 2--5 optical effective radii of the Sombrero galaxy (M\,104), which cannot be accounted for by the the cosmic X-ray background (CXB) and thus most likely reside in the halo of M\,104.
They postulated that these excess sources, detected at $L_{\rm X} \approx 10^{37-38}{\rm~erg~s^{-1}}$, may have a mixed origin. One possibility is LMXBs dynamically formed in globular clusters (i.e., GC-LMXBs), which tend to have a broader spatial distribution than the field stars. Another possibility is supernova-kicked LMXBs, which have been ejected from the inner galactic regions due to a strong recoil upon the supernova explosion that gives birth to the subsequently accreting NS or BH.   
\citet{Zhang2013} found a similar excess of halo X-ray sources in a sizable sample of mostly isolated ETGs. The significance of the excess was apparently correlated with both the GC specific frequency and the stellar mass of the host galaxy, supporting an origin of both GC-LMXBs and supernova-kicked LMXBs.

Using {\it Chandra} observations that sampled 80 intermediate-mass ETGs in the Virgo cluster, \citet{Hou2017ApJ} found that an excess of X-ray sources, with $L_{\rm X} \gtrsim 2\times 10^{38}{\rm~erg~s^{-1}}$, 
exist in the outskirt of these galaxies.
On the other hand, they found no significant excess sources of similar luminosities in a control sample of field ETGs. 
This strongly suggests that at least some of the excess sources are uniquely present
in the cluster environment and might be related to the so-called diffuse intra-cluster light (ICL; e.g., \citealp{Mihos2005ApJ,Mihos2017}).
In cluster/group environments, tidal interactions continue to shape and redistribute the stellar content of essentially all member galaxies. 
In particular, typically low-mass, infalling galaxies are stripped off of their stars, leading to the gradual building of the ICL, a generic process that can begin at the early stage of galaxy clustering \citep{Rudick2006,Rudick2011}. 
Tidal stripping can also facilitate the growth of an extended stellar halo around the brightest cluster galaxies (BCGs). 
Being predominantly old stellar populations, the ICL is expected to harbor X-ray sources such as LMXBs. 
In addition, the stripped relic of nucleated dwarf galaxies, typically of a compact appearance, may also be the host of bright X-ray sources (\citealp{Hou2016,Hou2017ApJ}). 
Conversely, a representative sample of intra-cluster X-ray sources can serve as a new tool to study the otherwise formidable ICL (\citealp{Finoguenov2002}), which conventionally requires sensitive optical observations due to its extremely low surface brightness.  

The Virgo, however, remains the only cluster in which intra-cluster X-ray sources have been probed and detected. 
The {\it Chandra} observations utilized by \citet{Hou2017ApJ} covered only a small portion ($\sim$3 deg$^{2}$) of Virgo, but its large angular size ($\sim$100 deg$^{2}$) renders a full mapping a challenging, if not infeasible, task with contemporary X-ray telescopes, thus limiting our ability to extend the search for intra-cluster X-ray sources in Virgo.
Located at a distance of $\sim$20.0~Mpc \citep{Blakeslee2009ApJ}, the Fornax cluster is a dynamically more evolved, more compact and less massive system compared to Virgo. 
Like Virgo, Fornax has been an important laboratory to explore the physics of hierarchical structure growth. 
Recent optical surveys including the {\it HST} ACS Fornax Cluster Survey (ACSFCS; \citealp{Jordan2007ApJS}) and the Fornax Deep Survey with VST (FDS; \citealp{Iodice2016ApJ}) have significantly advanced our knowledge about the ETGs \citep{Cote2007,Turner2012}, dwarf galaxies \citep{Venhola2017}, GC populations \citep{Jordan2015ApJS}, as well as the ICL \citep{Iodice2017ApJ} of this cluster.  
In the X-ray band, a survey of the inner $\sim$30$'$ region of Fornax has been conducted with {\it Chandra} observations \citep{Scharf2005ApJ}, which resulted in the detection of more than 700 point-like sources (including CXB sources) against an extended, asymmetric diffuse X-ray emission from the intra-cluster medium (\citealp{Ikebe1996,Jones1997}).
The X-ray sources in the BCG, NGC\,1399, have been extensively studied with a focus on their connection with GCs \citep{Angelini2001,Paolillo2011ApJ,DAgo2014}. 
The X-ray sources located outside the main stellar content of NGC\,1399, however, received little attention so far, except for the work of \citet{Phillipps2013MNRAS}, which studied the incidence rate of X-ray sources in compact stellar systems, including GCs and the so-called ultra-compact dwarfs (UCDs; \citealp{Phillipps2001}).

Using archival {\it Chandra} observations and assisted with recently advanced knowledge about the ICL in Fornax (see details below), here we search for the expected intra-cluster X-ray sources. The remainder of this paper is organized as follows. We introduce the {\it Chandra} observations and data reduction procedure in Section~\ref{sec:datas}.
Our procedures of X-ray source detection and characterization, along with the resultant source catalog, are described in Section~\ref{sec:detection}.
In Section~\ref{sec:analysis}, we provide strong evidence for the presence of intra-cluster X-ray sources by exploring the spatial and flux distributions of the detected sources. 
The possible origins of the intra-cluster X-ray sources are discussed in Section~\ref{sec:discussion}, followed by a summary in Section~\ref{sec:summary}.
Quoted errors throughout this work are at 68\% confidence level, unless otherwise noted.

\section{Data Preparation} \label{sec:datas}
The Fornax cluster, in particular its core region, has been extensively observed by {\it Chandra}, chiefly with its Advanced CCD Imaging Spectrometer (ACIS).
Initially we acquired a total of 30 archival {\it Chandra} observations, which include 12 ACIS-I and 18 ACIS-S observations.
Among these, ten ACIS-I observations were originally acquired by \citet{Scharf2005ApJ}, which provide a quasi-uniform exposure of $\sim$50 ks for the inner $\sim$30$'$ of Fornax.
Eight additional observations (6 ACIS-S and 2 ACIS-I) were pointed toward the BCG NGC\,1399, while another six ACIS-S observations focused on the infalling ETG NGC\,1404 \citep{Su2017}, which is centered at a projected radius of $\sim$9\farcm8 from the center of NGC\,1399.   
These deep exposures enable a good sensitivity for detecting point sources against the strong diffuse X-ray emission from the two galaxies.  

We reprocessed the archival level-1 data using CIAO v4.8 and the corresponding calibration files, following the standard procedure\footnote{http://cxc.harvard.edu/ciao/}.
We have examined the light curve of each observation and filtered time intervals that suffer from significant particle flares. 
In particular, ObsID 320 was found to be affected by a strong particle background during its entire 3.5-ks exposure and thus was completely discarded. 
The background filtering resulted in the final usage of 29 observations with a total cleaned exposure of $\sim$1300~ks. 
Across the combined field-of-view (FoV), the maximum exposure is $\sim$700 ks (on NGC\,1404) and the minimum is $\sim$40~ks. 
A log of the 29 observations is given in Table \ref{tab:journal}.

Following \citet{Hou2017ApJ}, we produced counts maps and exposure maps, at the natal pixel scale of 0\farcs492, for each ObsID in three energy bands: 0.5--2 ($S$-band), 2--8 ($H$-band) and 0.5--8 ($F$-band) keV. 
The exposure maps were weighted by an assumed incident spectrum of an absorbed power-law, with photon-index of 1.7 and absorption column density $N_{\rm H}=10^{21}\rm~cm^{-2}$.
This latter value is higher than the Galactic foreground absorption column ($\sim$$1.5\times10^{20}{\rm~cm^{-2}}$), but allows for some intrinsic absorption in LMXBs (e.g., \citealp{Luan2018}).
To ensure an optimal sensitivity for source detection, we included only data from CCDs I0, I1, I2 and I3 for the ACIS-I observations, and CCDs S2 and S3 for the ACIS-S observations. 
We also generated for each band and each ObsID maps of the point-spread function (PSF), at a given enclosed count fraction (ECF), using the same spectral weighting as for the exposure map.  
Lastly, we reprojected the individual counts maps or exposure maps to a common tangential point, i.e., the nucleus of NGC\,1399 ([R.A., Dec.]=[54.620941, -35.450657]), to produce a combined image. The PSF maps were similarly combined, with weights according to the local effective exposure. 

\begin{figure*}\centering
\includegraphics[scale=0.6,angle=0]{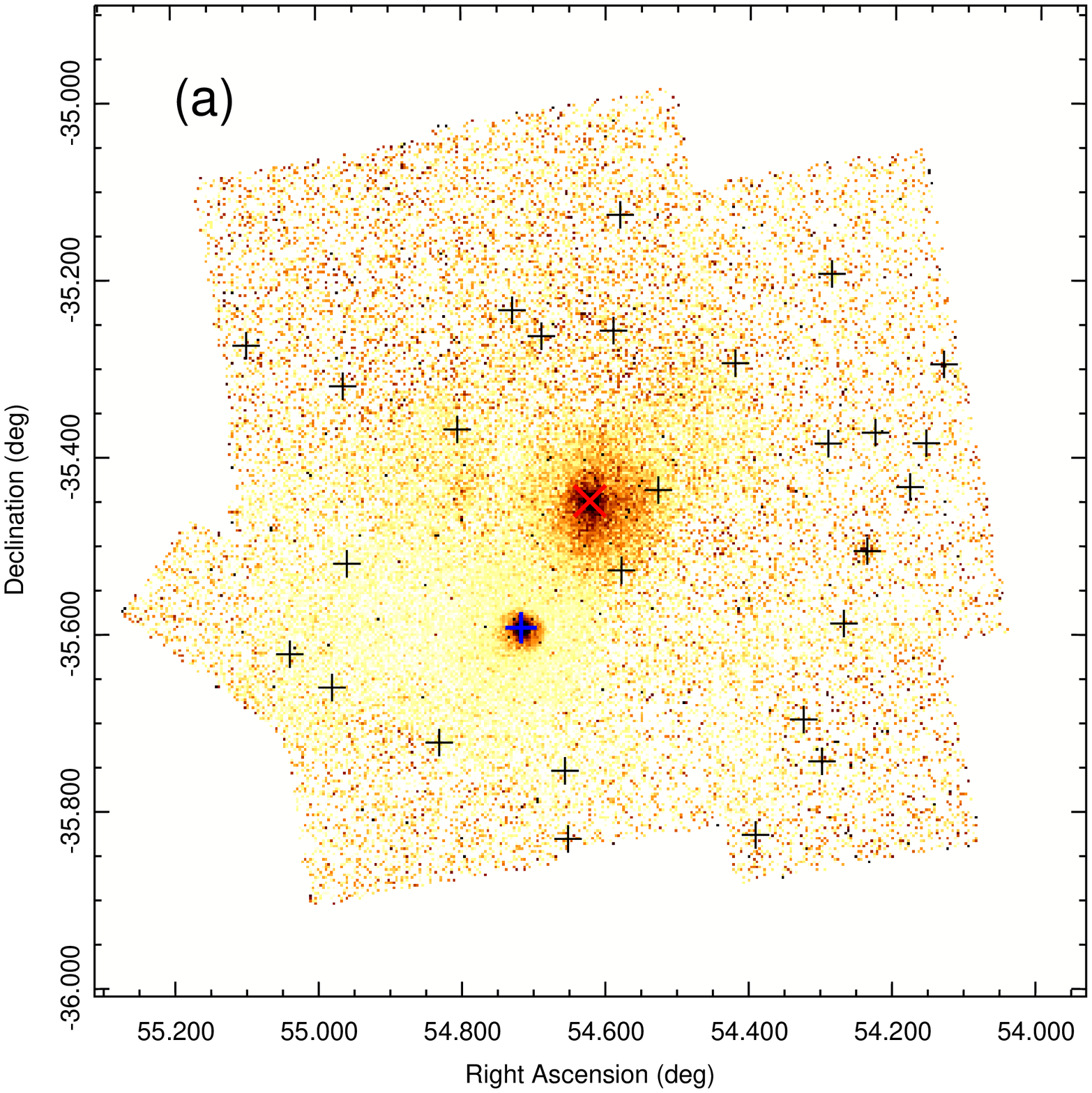}
\includegraphics[scale=0.3,angle=0]{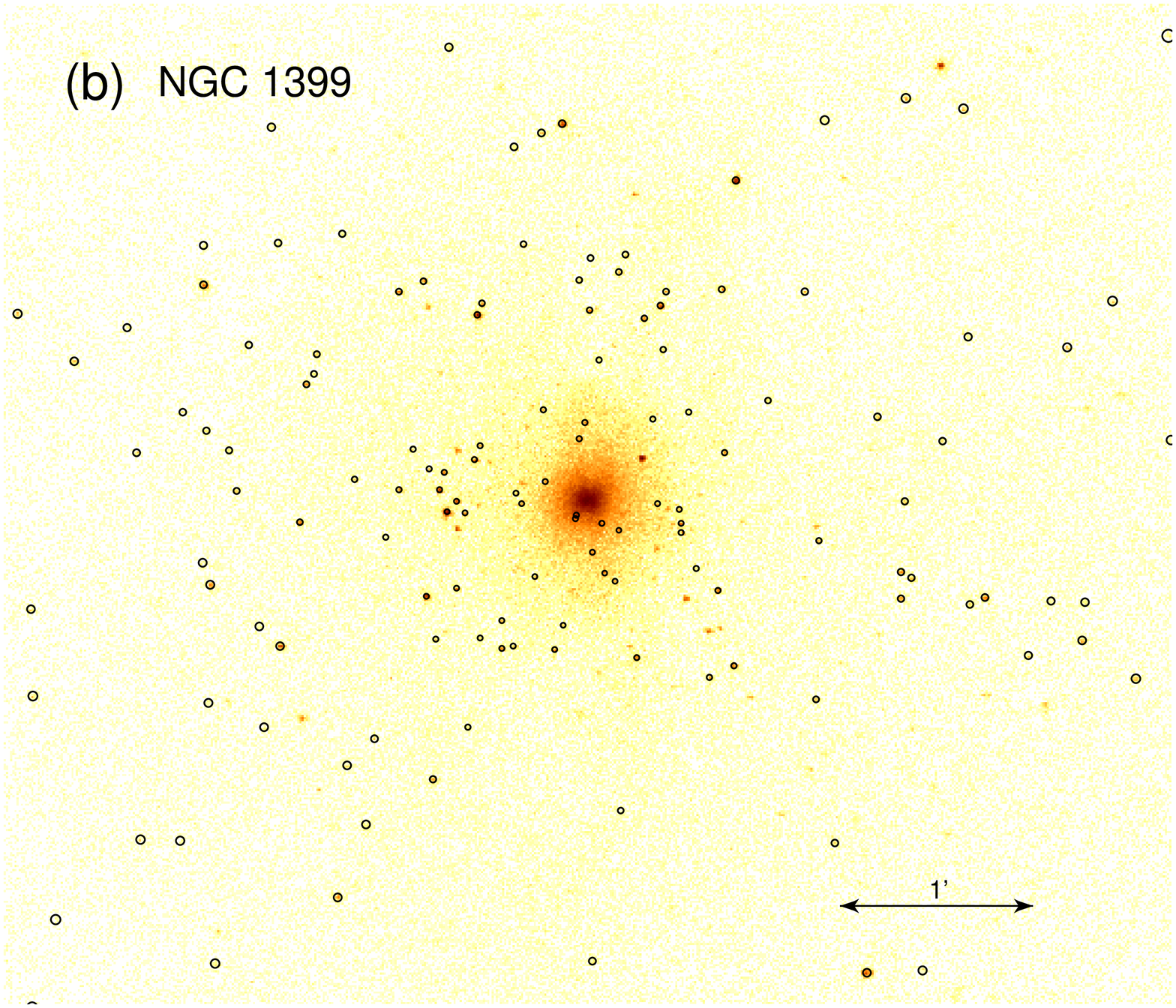}
\includegraphics[scale=0.3,angle=0]{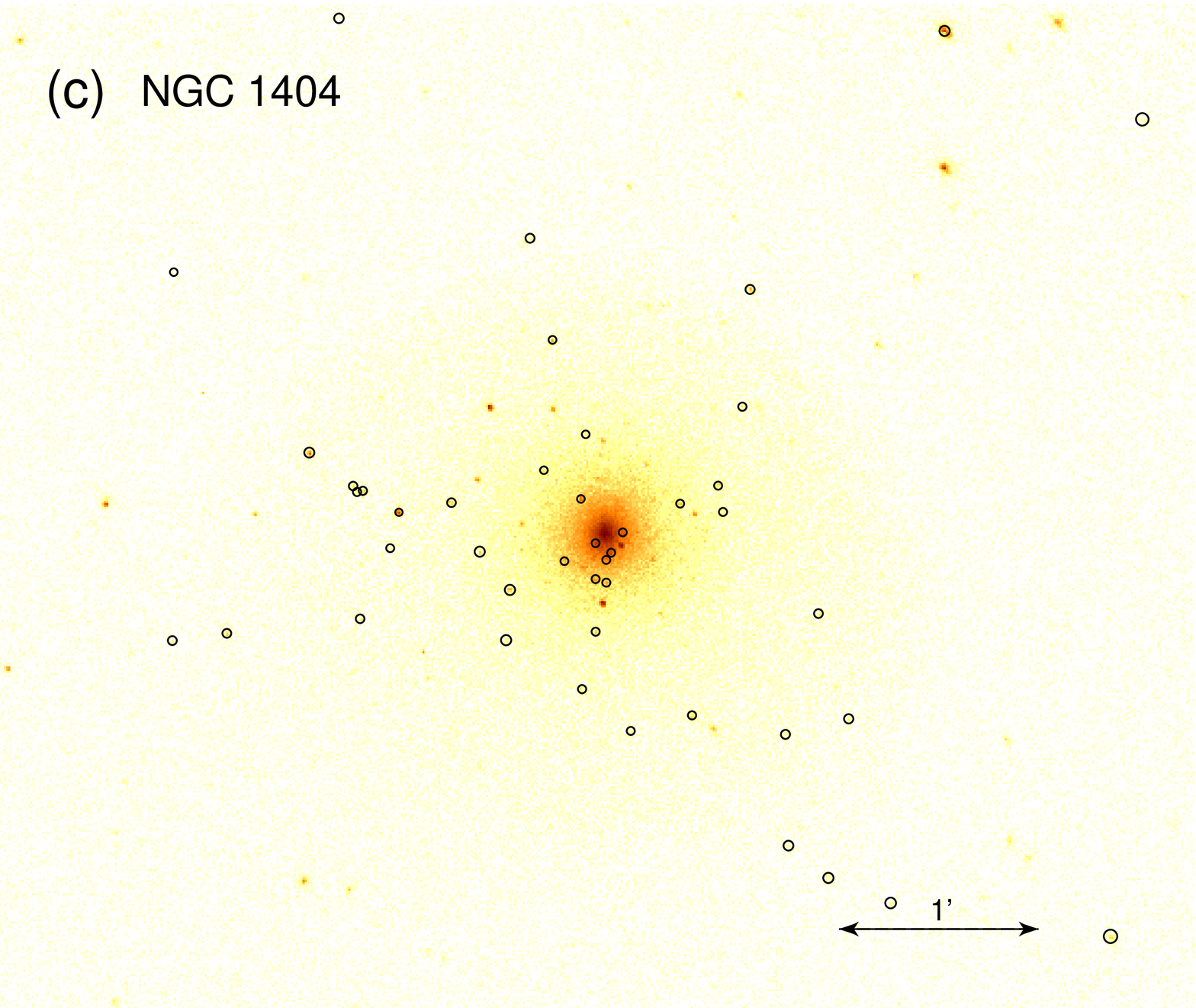}
\caption{(a) The mosaic 0.5--8 keV flux image of the Fornax cluster, smoothed by a 2-pixel Gaussian kernel. The center of NGC\,1399 and NGC\,1404 is marked by a red `X' and a blue `+', respectively. The locations of other FCC galaxies within the FoV (Table~\ref{tab:FCC}) are marked by black crosses. 
A close-up view of the vincity of NGC\,1399 and NGC\,1404 is shown in panel (b) and (c), respectively, where deep exposures are available. 
The small black circles highlight X-ray sources associated with known globular clusters. See Section~\ref{sec:detection} for details.
}
\label{fig:exp}
\end{figure*}

Figure~\ref{fig:exp}a presents the 0.5--8 keV flux image of the combined FoV. 
A close-up view of the vicinity of NGC\,1399 and NGC\,1404, where both strong diffuse X-ray emission and numerous discrete sources are clearly present, is displayed in Figure~\ref{fig:exp}b and \ref{fig:exp}c, respectively.

\section{X-ray Source detection} \label{sec:detection}
Our source detection and source characterization procedures are as detailed in \citet{Zhu2018}. Here we briefly outline the key steps and results. 

i) We performed source detection for each of the three energy bands over the combined FoV, using the CIAO tool {\it wavdetect}. The combined exposure map and 50\%-ECF PSF map were supplied to the detection process.
A false-positive probability threshold of $10^{-6}$ was adopted. 
This resulted in a raw list of 1008 sources in $S$-band, 758 sources in $H$-band, and 1248 sources in $F$-band. 
At this point, we also produced a sensitivity map, which records the detection limit at each pixel, following the recipe of \citet{Kashyap2010ApJ} and according to the chosen false-positive threshold. 

ii)  We refined the source centroids output from {\it wavdetect}, using a maximum likelihood method \citep{Boese2001} that iterates over the recorded positions of the individual counts within the 90\% enclosed counts radius (ECR) as default. 
A 50\% ECR was adopted for a small fraction of sources to avoid confusion with closely neighboring sources. 
The position uncertainty ($PU$) at 68\% confidence level was estimated following the empirical relation between $PU$, source counts and source position in terms of the off-axis angle \citep{Kim2007ApJS}.

iii) We then performed source photometry to derive the exposure- and PSF-corrected photon flux. 
A circular aperture was chosen, with a default 90\% ECR to extract the source counts; for those sources subject to crowding, the 50\% ECR was again adopted.  
The background extraction regions were typically 2--4 times the 90\% ECR, excluding pixels falling within 2 times the 90\% ECR of any neighboring sources.
A Bayesian approach was employed to calculate the photon fluxes and bounds, which takes care of the Poisson statistics at the low-count regime \citep{Park2006ApJ}.
We also calculated the hardness ratio, defined as $HR = (S_{\rm 2-8}-S_{\rm 0.5-2})/(S_{\rm 2-8}+S_{\rm 0.5-2})$, where $S_{\rm 0.5-2}$ and $S_{\rm 0.5-2}$ are the photon flux of the $S$-band and $H$-band, respectively.

iv) We calculated the binomial no-source probability ($P_{\rm B}$; \citealp{Weisskopf2007ApJ}) to filter spurious sources due to background fluctuation.
Any source with $P_B > 0.01$ is considered spurious and excluded from further analysis, while the remaining are considered genuine point sources.



v) A cross-matching method \citep{Hong2009ApJ} was then employed to identify the same source detected in more than one bands.
The relative distance ($d_r$) of two sources from two different bands, is defined as the ratio of the angular offset between the source centroids to the quadratic sum of the respective $PU$. 
We required $d_r < 3.0$ for a true source pair. 
With this criterion, we estimate 8 (7) random matches out of the 876 (659) $S/F$ ($H/F$) pairs, i.e., $\sim$1\% false matches. 

At this point, we arrive at a list of 1279 independent sources, among which 
1177 are detected in the $F$-band, 924 in the $S$-band and 713 in the $H$-band.
For future reference, we present in Table~\ref{tab:sourcecatalog} a catalog of basic source properties, including centroid position, observed photon flux in the three bands, 0.5--8 keV unabsorbed energy flux and the hardness ratio. 
To derive the 0.5--8 keV unabsorbed flux, we have adopted a photon flux-to-energy flux conversion factor of $3.64\times10^{-9}{\rm~erg~ph^{-1}}$ according to the assumed incident source spectrum (Section~\ref{sec:datas}). 

\begin{figure*}\centering
\includegraphics[width=\textwidth]{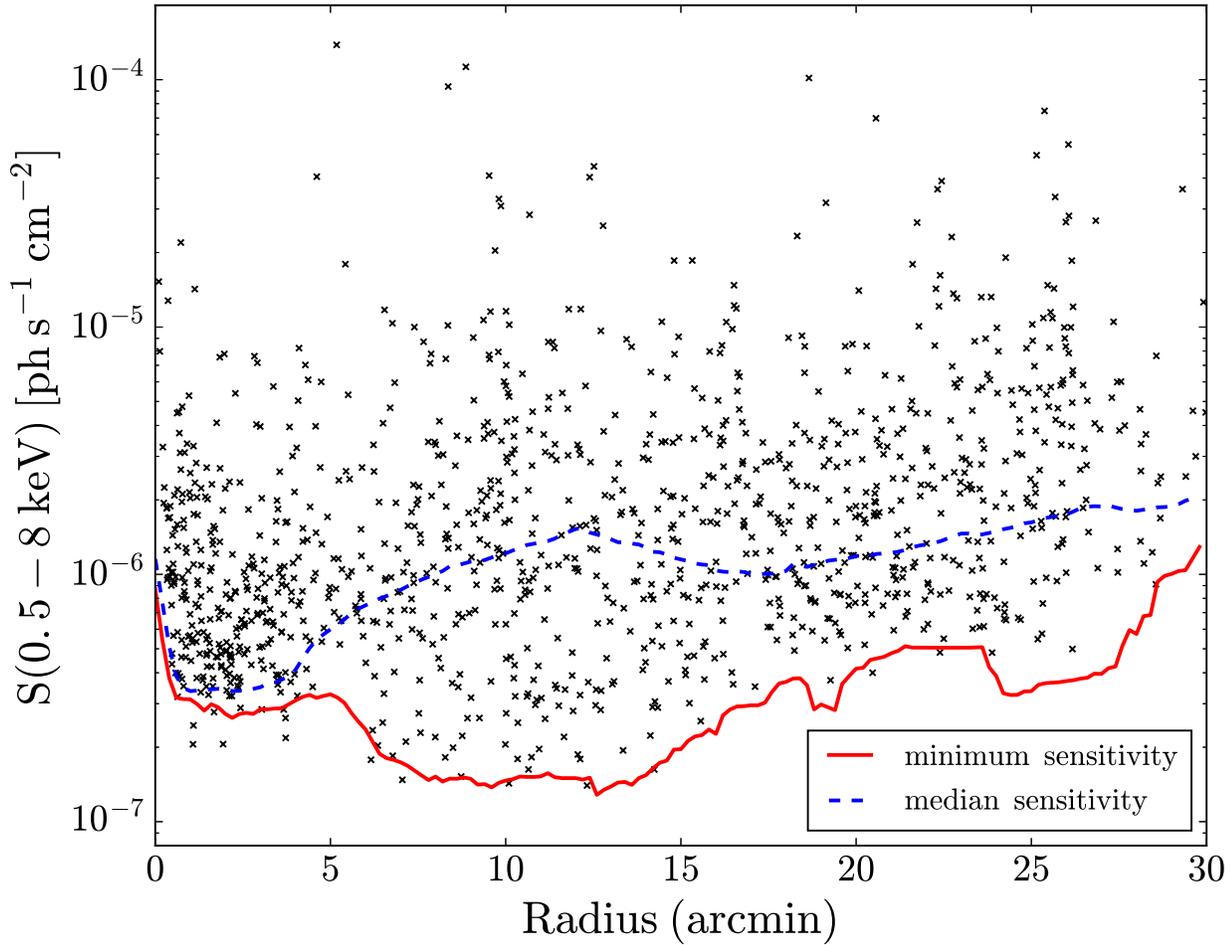}
\caption{Observed 0.5--8 keV photon flux versus the projected distance from the center of NGC\,1399, for sources detected in the $F$-band.
The red solid and blue dashed curves denote the azimuthally minimum and median detection sensitivity, respectively. 
The ``valley" between 6{\arcmin}--13{\arcmin} in the minimum sensitivity is due to the deep exposure towards the vicinity of NGC\,1404.}
\label{fig:pfvsr}
\end{figure*}

\begin{figure*}\centering
\includegraphics[width=\textwidth]{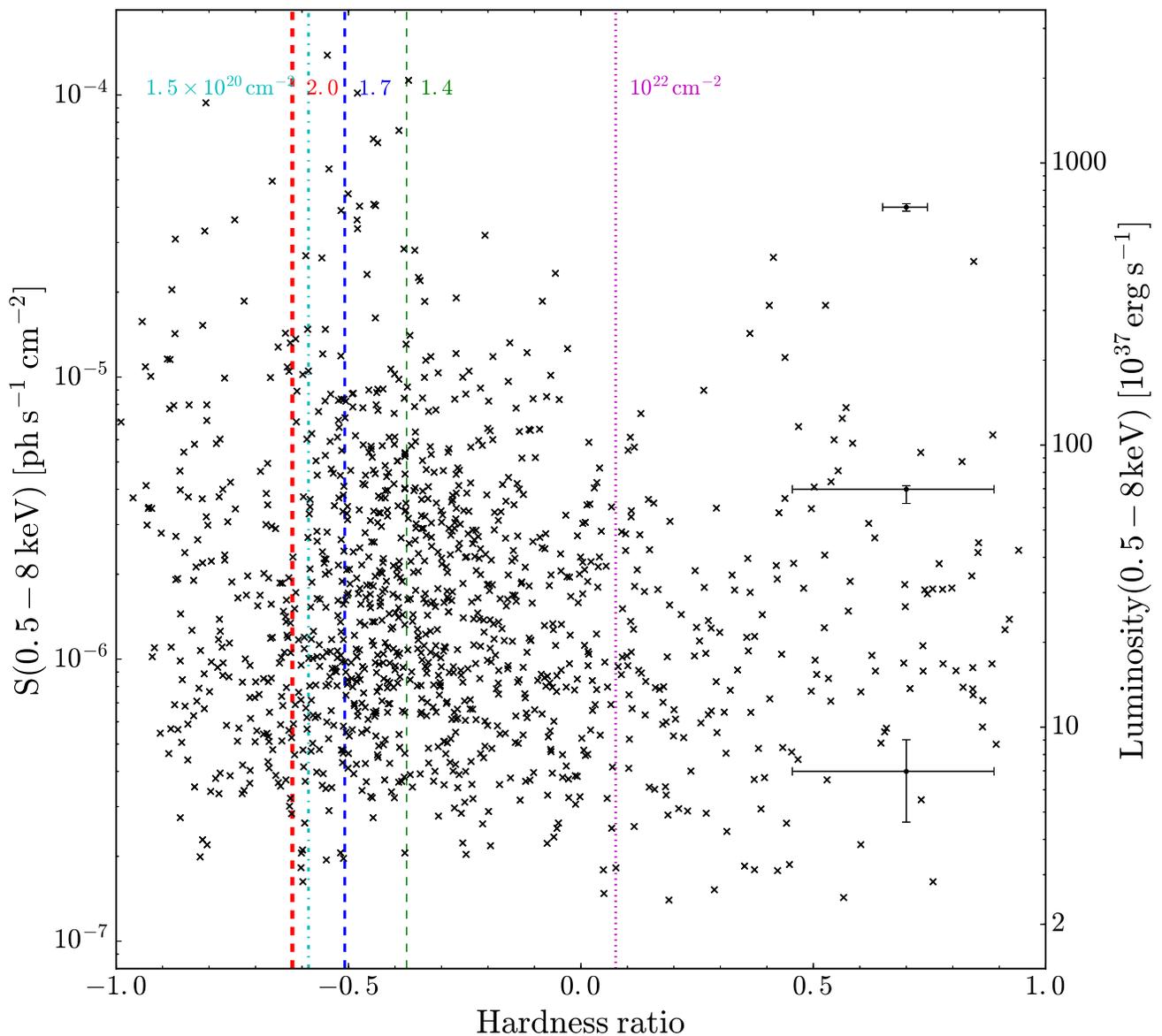}
\caption{Hardness ratio versus observed 0.5--8 keV photon flux, for sources detected in the $F$-band. The y-axis on the right shows the corresponding unabsorbed luminosity assuming the distance of Fornax. The vertical dashed lines represent predicted hardness ratios by various absorbed power-law spectral models. The cyan dash-dotted line represents column density $N_{\rm H} = 1.5\times10^{20}{\rm~cm^{-2}}$ and photon-index of 1.7; the red, blue and green dash lines represent $N_{\rm H} = 10^{21}\;\!{\rm cm}^{-2}$ and photon-index of 2.0, 1.7 and 1.4, respectively; the magenta dotted line represents $N_{\rm H} = 10^{22}\;\!{\rm cm}^{-2}$ and photon-index of 1.7. The three error bars, from top to bottom, indicate the median errors of sources with $S_{0.5-8} > 10^{-5}\;\! {\rm ph\;\!cm^{-2}\;\!s^{-1}}$, $10^{-6}\;\! {\rm ph\;\!cm^{-2}\;\!s^{-1}} < S_{0.5-8} < 10^{-5}\;\! {\rm ph\;\!cm^{-2}\;\!s^{-1}}$, and $S_{0.5-8} < 10^{-6}\;\! {\rm ph\;\!cm^{-2}\;\!s^{-1}}$.}
\label{fig:hr}
\end{figure*}

In Figure~\ref{fig:pfvsr}, we plot the observed 0.5--8 keV photon flux ($S_{0.5-8}$) versus the projected distance ($R$) from the cluster center (here defined as the center of NGC\,1399) for the $F$-band sources. 
For comparison, we also show the azimuthally-averaged detection limit as a function of $R$, which is derived from the sensitivity map. 
A limiting photon flux of $\sim$$3\times10^{-7}{\rm~ph~cm^{-2}~s^{-1}}$ is reached within $R \approx 4'$, where a deep exposure competes with the strong diffuse emission from NGC\,1399. 
It is noteworthy that the global limiting flux, $\sim$$1.5\times10^{-7}{\rm~ph~cm^{-2}~s^{-1}}$, is actually achieved around NGC\,1404. 
Outside NGC\,1399 and NGC\,1404, and out to $R\approx25'$, the median detection limit is roughly leveled at (1--2)$\times10^{-6}{\rm~ph~cm^{-2}~s^{-1}}$ due to the highly uniform effective exposure. 
Figure~\ref{fig:hr} shows $HR$ versus $S_{0.5-8}$ for the $F$-band sources, along with predicted hardness ratios of certain absorbed power-laws. It can be seen that the majority of sources exhibit $HR$ values consistent with them being LMXBs and/or background AGNs, i.e., with photon-indices of 1--2. 
A small number of hard sources ($HR \gtrsim 0.5$) might be heavily obscured AGNs.  

\subsection{Optical counterpart} {\label{subsec:counterpart}}
Our last step before turning to search for intra-cluster X-ray sources, involves the identification of {\it a priori} irrelevant sources on a best-effort basis. 
First, we identify possible Galactic foreground X-ray sources by cross-correlating with the USNO-B catalog \citep{Monet2003AJ} for optical sources with significant proper motions. This results in 18 pairs for a matching radius of {1\arcsec}.
Second, we identify any source that is located within 3{\arcsec} from the optical nucleus of any member galaxy of Fornax. 
Within the {\it Chandra} FoV, there are 29 known member galaxies according to the Fornax Cluster Catalog (FCC; \citealp{Ferguson1989AJ}), whose positions are marked in Figure~\ref{fig:exp}a.
Among them, a nuclear X-ray source is found in NGC\,1381 and NGC\,1387; in each case of NGC\,1399 and NGC\,1404, two sources satisfy our matching criterion, and the one with the smaller offset is chosen as the nuclear source. 
The basic information of the 29 FCC galaxies is summarized in Table \ref{tab:FCC}.

In addition, we identify X-ray sources positionally coincident with known GCs in Fornax. 
First, we consult with the ACSFCS GC catalog \citep{Jordan2015ApJS}, which, with a limiting $g$-band magnitude of 26.3, is expected to contain $\gtrsim$90\% of the GC population that falls within the ACS fields. 
We select GC candidates with $p_{\rm GC}\geq0.5$ in the ACSFCS catalog and adopt a matching radius of 1\arcsec\ between the X-ray and optical centroids, which results in 134 pairs.
Since the ACSFCS fields only cover a small fraction of the {\it Chandra} FoV, we also incorporate the GC catalog based on the FDS (\citealp{Cantiello2018A&A}),  
which fully overlaps with the {\it Chandra} FoV and reaches a $g$-band limiting magnitude of $\sim$24.0.
A matching radius of 1{\arcsec} results in 74 pairs from the FDS catalog.
We then cross-correlate two additional GC catalogs, which are intermediate between the ACSFCS and FDS catalogs in terms of sky coverage and sensitivity.
A matching radius of 1\arcsec\ results in 118 pairs from the catalog of \citet{Kim2013ApJ}, which is based on Blanco-4m observations that cover a 36'x36' field around NGC\,1399 down to a limiting $U$-band magnitude of 24.4; 
the same matching radius finds 130 pairs from the catalog of \citet{Paolillo2011ApJ}, which is based on ACS $V$-band mapping of a $\sim$$10'\times10'$ field centered at NGC\,1399. 

In total, we have identified 270 independent X-ray sources associated with Fornax GCs, which are presumably GC-LMXBs.   
By artificially shifting the X-ray centroids by 10{\arcsec} in both directions of R.A. and Dec., we estimate that overall only a few percent of all GC-LMXB pairs can be random matches, although the amount increases to $\sim$20\% in the inner few arc-minutes of NGC\,1399, where the surface density of both X-ray and optical sources is high. 
On the other hand, due to the heterogeneous nature of the GC catalogs in use, it is not straightforward to assess the completeness of GC-LMXBs. 
Empirically, the vast majority of GC-LMXBs are found in massive GCs (e.g., \citealp{Li2010ApJ,Hou2016}). 
Indeed, among the 134 ACSFCS GC-LMXB pairs, $\sim$73\% are found in GCs brighter than the so-called turnover magnitude ($\sim$24.0 mag of $g$-band at the distance of Fornax; \citealp{Villegas2010ApJ}). 
We can have a rough estimate of potentially missing GC-LMXBs in the other catalogs, by scaling the number of pairs found at a certain magnitude range with respect to the ACSFCS catalog. 
In particular, the FDS catalog, which covers the entire {\it Chandra} FoV and just reaches the turnover magnitude, is expected to miss $\lesssim$20 fainter, low-mass GCs that might have an X-ray counterpart. 

We further cross-correlate the detected X-ray sources with catalogs of Fornax UCDs \citep{Gregg2009AJ,Voggel2016A&A}, finding 15 pairs with a matching radius of 1{\arcsec}.
It turns out that among them, 13 have already been classified as a GC. This is not totally surprisingly, since some UCDs can appear as ``giant" GCs in optical images. 
The other two UCDs (source ID 326 \& 717 in Table~\ref{tab:sourcecatalog}) are new identifications.


The above X-ray sources with an identified optical counterpart is denoted in the last column of Table~\ref{tab:sourcecatalog} (`F' for foreground sources, `N' for nuclear sources, `G' for GC-LMXBs, and `U' for UCDs).

\section{Analysis: Probing Intra-cluster X-ray sources} \label{sec:analysis}
\subsection{Spatial Distribution} \label{sec:spatial}

The presence of intra-cluster X-ray sources, defined as sources that are located in Fornax but outside the main stellar content of the member galaxies, is best probed by the source spatial distribution \citep{Hou2017ApJ}.
We focus on the sources detected in 0.5--8 keV band, having excluded foreground sources, nuclear sources, UCDs and probable GC-LXMBs as identified in the previous section. 
It is noteworthy that some GC-LMXBs can belong to the putative intra-cluster stellar populations. 
Indeed, in the case of Virgo, $\sim$30\% of the ``excess" X-ray sources can be attributed to GC-LMXBs \citep{Hou2017ApJ}.
Here we preclude known GC-LMXBs (Section~\ref{subsec:counterpart}) to focus on the ``field" population. The relevance of GC-LMXBs to the intra-cluster populations will be further addressed in Section~\ref{sec:discussion}.

\begin{figure*}\centering
\includegraphics[scale=0.42,angle=0]{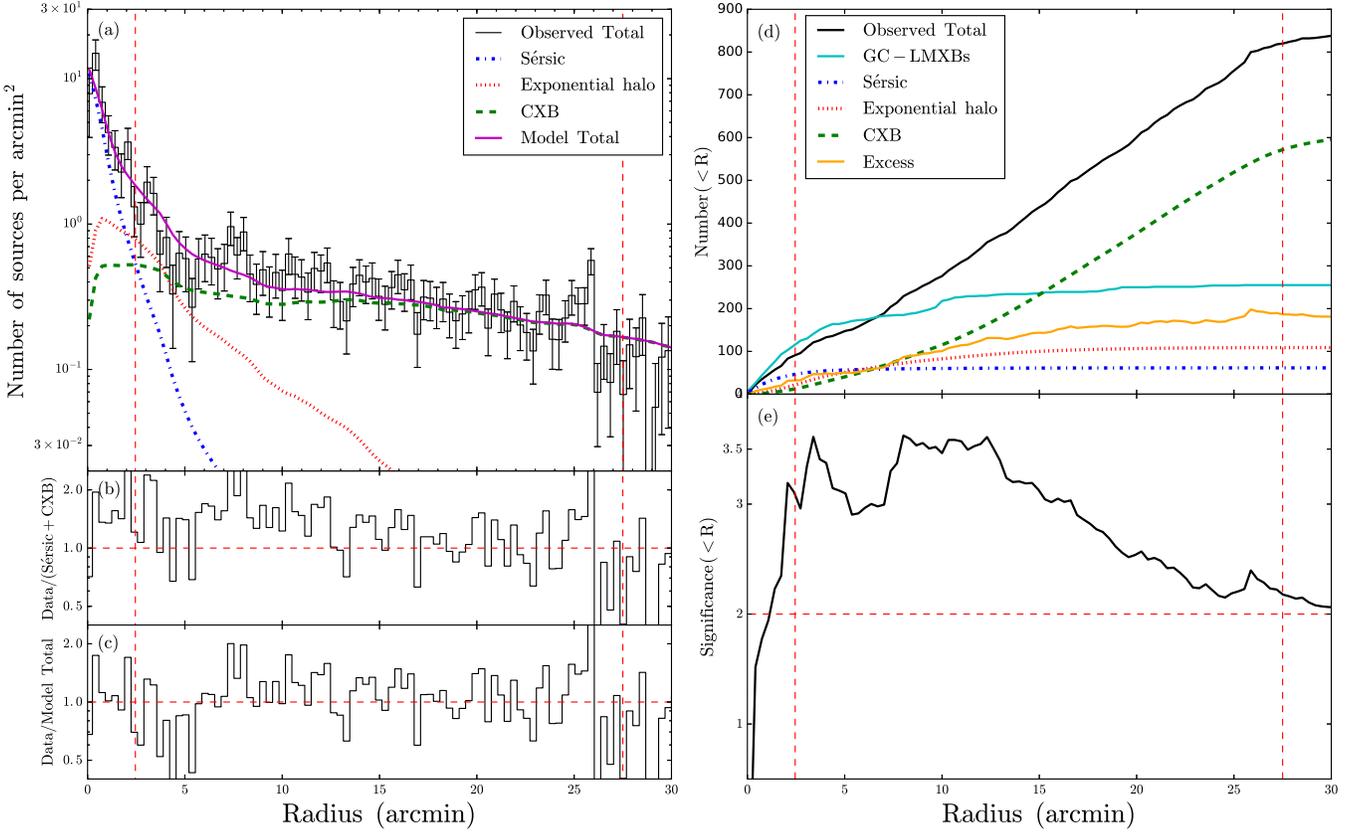}
\caption
{(a): The surface number density of sources detected in the $F$-band (black data points with 1\,$\sigma$ error bars), excluding foreground stars, nuclear sources, sources associated with member galaxies (except NGC~1399) and GC-LMXBs. 
The green dashed curve is the predicted profile of CXB sources;
the blue dot-dashed curve is the fitted S{\'e}rsic profile for field sources of NGC\,1399; the red dotted curve is the fitted exponential profile for sources associated with the extended halo. 
The magenta solid curve is the sum of CXB, S{\'e}rsic and exponential components. 
All components are modified by detection incompleteness. 
(b) Ratio between the observed and modeled source surface densities, where the model includes only the CXB and S{\'e}rsic components. 
(c) Ratio between the observed and modeled source surface densities, where the model includes all three components.  
(d) Cumulative number of observed and predicted sources. The various source components are as denoted by the insert. Excess refers to $N_{\rm excess} = N_{\rm obs}-N_{\rm CXB}-N_{\rm ser}$.
(e) The cumulative significance of excess sources, as defined by Eqn.~2.
In all panels, the inner vertical line marks 3 times the effective radius of NGC\,1399, while the outer vertical line marks the radius beyond which the  azimuthal coverage of the {\it Chandra} FoV drops below 50\%. See text for details. 
}
\label{fig:rp}
\end{figure*}

In addition, we preclude sources that are most likely associated with the member galaxies other than NGC\,1399. 
Such sources are defined as those falling within three times the optical effective radius ($R_{\rm e}$) of a given host galaxy (Table 3), a practical border of its bulk stellar content. 
This leads to the further removal of 51 sources from the subsequent analysis. 
The cumulative number of the remaining sources (849 in total) as a function of projected distance from the center of NGC\,1399 (also the cluster center) is plotted as the black solid curve in Figure~\ref{fig:rp}d.
The surface density profile of these sources, corrected for the masked area around the 28 member galaxies, is shown as the black histogram in Figure~\ref{fig:rp}a.  

By its construction, the surface density profile should consist of at least two main components: (i) an inward rising component related to the field stars of NGC\,1399, and (ii) an intrinsically flat component arising from the CXB, which dominates the profile at large radii. Both trends are clearly evident in Figure~\ref{fig:rp}a. 
We characterize these two components using empirical models, following the method of \citet{Hou2017ApJ}.

The CXB contribution is estimated from the 0.5--10 keV $\log{N}$--$\log{S}$ relation of \citet{Georgakakis2008MNRAS}, 
for which we have converted the $F$-band photon flux into the intrinsic 0.5--10~keV energy flux, assuming a fiducial absorbed power-law spectrum with photon-index of 1.4 and $N_{\rm H}=1.5\times10^{20}\;\!{\rm cm^{-2}}$.
The predicted radial distribution of the CXB, corrected for detection incompleteness according to the sensitivity map, is plotted as the green dashed curve in Figure~\ref{fig:rp}a. It is obvious that the observed and predicted source profiles are highly consistent with each other at $R \gtrsim 20'$.

The central component, presumably field-LMXBs, is expected to follow the starlight distribution of NGC\,1399, which can be characterized by a S{\'e}rsic law, 
\begin{equation}\centering
\begin{aligned}
\mu(R)=\mu_{\rm e}+k_n\;\![(\frac{R}{R_{\rm e}})^{\frac{1}{n}}-1],
\end{aligned}
\end{equation}
where the parameters were determined from the FDS \citep{Iodice2016ApJ}: $\mu_e=21.5\;{\rm mag\;arcsec^{-2}}$, $R_{\rm e}=49.1\;{\rm arcsec}$, $n=4.5$ ($k_n=2.17\;\!n-0.355$; \citealp{Caon1993MNRAS}) in the $g$-band.
We note that these parameters were derived by excluding the stellar core ($R<0\farcm1$) of NGC\,1399, hence we also neglect the few X-ray sources at $R<0\farcm1$ in the following comparison.
To convert the starlight into stellar mass, we adopt the color-dependent mass-to-light ratio of \citet{Bell2003ApJS}, 
finding $M/L \approx 5.2$ for $g$-band Solar absolute magnitude of 5.15 and a quasi-uniform color $g-i=1.2$ \citep{Iodice2016ApJ}. 
The S{\'e}rsic profile is then normalized to match the observed X-ray source profile within $R \lesssim 3R_{\rm e} \approx 2\farcm5 $ (blue dotted curve in Figure~\ref{fig:rp}a). 
We have applied the field-LMXB luminosity function (LF) of \citet{Zhang2011A&A} to correct the S{\'e}rsic profile for the position-dependent detection incompleteness. 
The normalization ($\epsilon_{\rm X}$) of the S{\'e}rsic profile is found to be $\sim$4.3 sources per $10^{10}{\rm ~M_\odot}$ (above a reference luminosity $L_{\rm ref}=5\times10^{37}{\rm~erg~s^{-1}}$). 
This value is in good agreement with the range of specific number of LMXBs above a similar limiting luminosity found in ETGs (2.8--5.4 sources per $10^{10}{\rm ~M_\odot}$, \citealp{Zhang2011A&A, Zhang2012}).

The combination of the above two components, however, cannot fully account for the observed source profile; an excess is clearly seen over $3' \lesssim R \lesssim 15'$ (Figure~\ref{fig:rp}b). 
We stress that adjusting $\epsilon_{\rm X}$, the only free parameter, does not eliminate the excess, due to the steeply declining S{\'e}rsic profile.
{\it We define the excess as $N_{\rm excess} = N_{\rm obs}-N_{\rm ser}-N_{\rm CXB}$, and quantify its cumulative significance as:}
\begin{equation}
\begin{aligned}
{\rm SIG}=\frac{N_{\rm obs}-N_{\rm ser}-N_{\rm CXB}}{\sqrt{N^2_{\rm obs}\sigma^2_{\rm P}+N^2_{\rm CXB}\sigma^2_{\rm c}}},
\end{aligned}
\end{equation}
where $N_{\rm obs}$ is the number of observed sources, $N_{\rm ser}$ and $N_{\rm CXB}$ are the number of predicted sources in the S{\'e}rsic and CXB components, respectively. 
We take into account the Poisson variance $\sigma^2_{\rm P}=1/{N_{\rm obs}}$ and the cosmic variance $\sigma^2_{\rm c}$, but neglects the uncertainty in $N_{\rm ser}$ which is small compared to the other two errors over the radial range where the excess is concerned. 
The cosmic variance can be estimated as \citep{Lahav1992ApJ},
\begin{equation}
\begin{aligned}
\sigma^2_{\rm c}=\frac{1}{\Omega^2}\int \omega(\theta)\;\!{\rm d}\Omega_1\;\!{\rm d}\Omega_2=C_{\gamma}\;\!\theta_0^{\gamma-1}\;\!\Theta^{1-\gamma}
\end{aligned}
\end{equation}
where $\omega(\theta)=(\theta/\theta_0)^{1-\gamma}$ is a power-law angular correlation function \citep{1980lssu.book}, $\theta_0 \approx 0.00214$ deg is the correlation length \citep{Ebrero2009A&A}, $\Omega=\Theta^2 \approx 0.66 {\rm~deg^2}$ is the size of the FoV, and the numerical factor $C_{\gamma} \approx 2.25$ for the canonical value of $\gamma=1.8$ \citep{1980lssu.book}. 
This results in $\sigma_{\rm c} \approx 0.14$.
As shown in Figure~\ref{fig:rp}e, the excess has a cumulative significance $\geq2\,\sigma$ everywhere beyond $R = 3\,R_{\rm e}$, reaching a maximum of $3.6\,\sigma$ at $7\farcm5 \lesssim R \lesssim 12\farcm5$. 

Based on the FDS, \citet{Iodice2016ApJ} detected an extended and diffuse stellar halo around NGC\,1399, which is distinct from the S{\'e}rsic component and can be traced out to a projected distance of $\sim$190 kpc. The azimuthally-averaged surface brightness profile of this halo was described by an exponential law: 
\begin{equation}\centering
\begin{aligned}
\mu(R)=\mu_0+1.086({R}/{R_{\rm h}}),
\end{aligned}
\end{equation}
where in the $g$-band $\mu_0=23.4\;{\rm mag\;arcsec^{-2}}$ and $R_{\rm h}=292\;{\rm arcsec}$. 
It is conceivable that the at least part of the excess X-ray sources can be attributed to this stellar halo. 
We test this possibility by adding the above exponential component to our spatial model, again correcting for detection incompleteness with respect to the empirical LF of field-LMXBs and adopting the same $M/L$ as for the S{\'e}rsic component. 
We then fit the observed source profile using the S{\'e}rsic+exponential+CXB model over the range of $0\farcm1 < R < 27\farcm5$, the outer boundary chosen as the radius beyond which the azimuthal coverage of the {\it Chandra} FoV drops below 50\%. 
Initially we assume that the same normalization applies to both the S{\'e}rsic and exponential components, finding $\epsilon_{\rm X} = 4.3\pm0.4$ sources per $10^{10}{\rm ~M_\odot}$ (above $L_{\rm ref}$) and $\chi^2/{\rm d.o.f.}=64.2/54$.
The fitted exponential component, with a total number of $N_{\rm exp} = 109$ sources, is plotted as the red dotted curve in Figure~\ref{fig:rp}a, while the combined model is plotted as the solid curve. 
This simple exercise suggests that the exponential halo can be responsible for a substantial fraction of the excess sources.
On the other hand, as indicated in Figure~\ref{fig:rp}c,d, even after accounting for the exponential component, there is still a significant residual (i.e., $N_{\rm obs}-N_{\rm ser}-N_{\rm exp}-N_{\rm CXB}$) of $\sim$74 sources (or 40\% of all excess sources).
We also test the case that the two spatial components have different normalizations, obtaining $\epsilon_{\rm X}$ (S{\'e}rsic) = $5.3\pm1.0$ sources per $10^{10}{\rm ~M_\odot}$ and $\epsilon_{\rm X}$ (exponential) = $3.5\pm0.9$ sources per $10^{10}{\rm ~M_\odot}$, with $\chi^2/{\rm d.o.f.}=62.9/53$. This indicates no significant difference in the source abundance between the two components. 
The role of the stellar halo of NGC\,1399 in the excess sources will be further addressed in Section 5.

\subsection{Luminosity Function} \label{sec:lf}
The source properties can be further constrained by examining their LF. 
In the case of Virgo, however, the moderate number and limited flux range of the excess sources did not permit a meaningful LF analysis \citep{Hou2017ApJ}. 
Here we consider three sets of $F$-band sources: (i) `Fornax-field' refers to sources detected between $4\arcmin < R < 15\arcmin$, excluding known GC-LMXBs and UCDs but necessarily including CXB contribution; (ii) `NGC\,1399-field' refers to field sources detected at $R < 3R_{\rm e}$, where CXB contribution is negligible (Figure~\ref{fig:rp}),  and (iii) `Fornax-GC' refers to GC-LMXBs detected between $4\arcmin < R < 15\arcmin$, but excluding 4 objects that are also classified a UCD\footnote{According to \citet{Hou2016}, the X-ray emission from UCDs is most likely due to dynamically-formed LMXBs, as in the case of GCs. Here we conservatively distinguish the X-ray counterparts of UCDs and GCs.}.

\begin{figure*} \centering
\includegraphics[scale=0.8,angle=0]{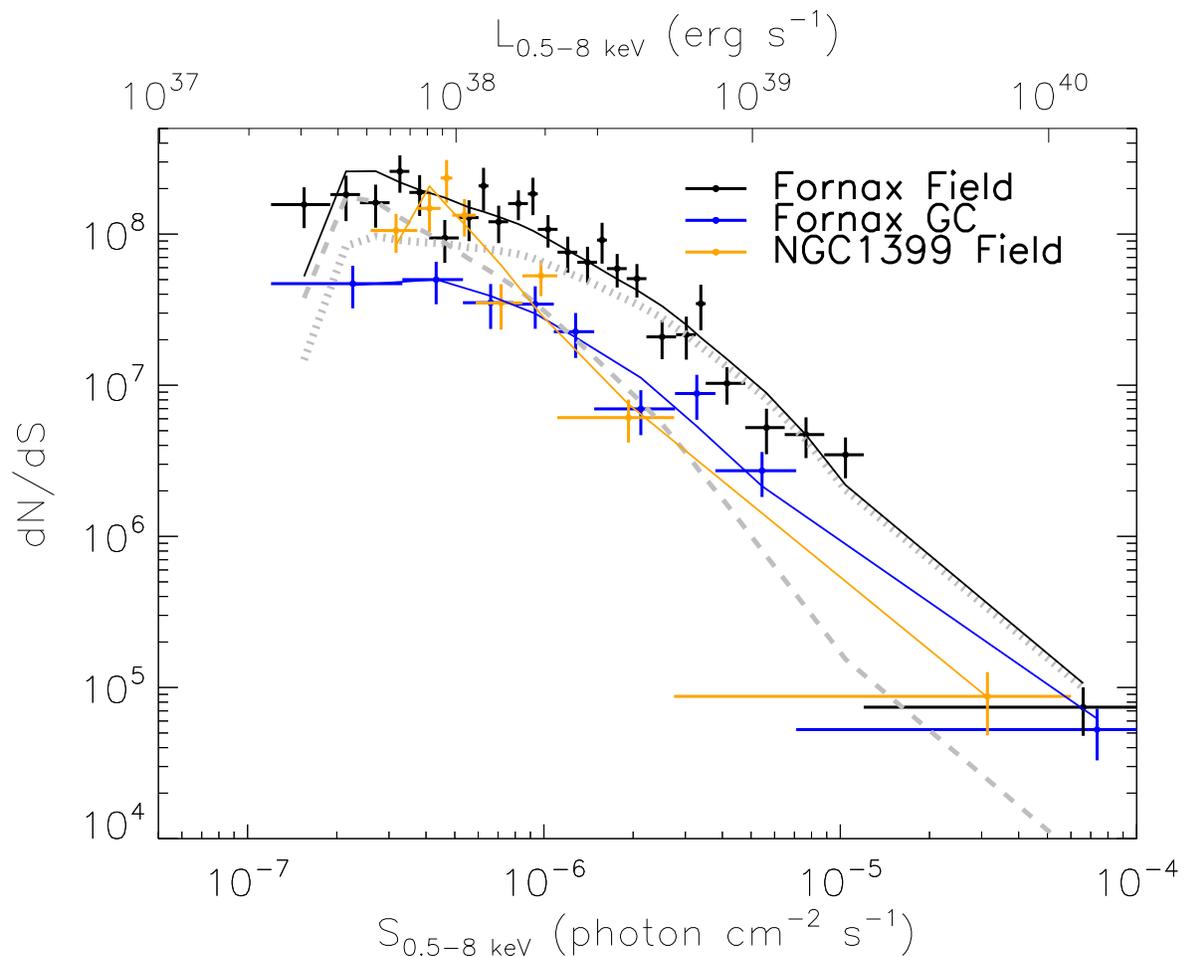}
\caption
{The observed luminosity function of $F$-band sources, adaptively binned to have a minimum of 9 sources per flux bin. 
Black (Fornax-field): field sources (i.e., excluding GC-LMXBs and sources associated with member galaxies) detected at $4\arcmin < R < 15\arcmin$. 
Blue (Fornax-GC): GC-LMXBs detected at $4\arcmin < R < 15\arcmin$.
Orange (NGC\,1399-field): field sources detected at $R < 3\,R_{\rm e}$.
The fitted model, modified by detection incompleteness and Eddington bias, is plotted as the solid curve. For Fornax-field, the model includes a power-law (dashed curve) and a fixed CXB component (dotted curve). For NGC\,1399-field and Fornax-GC, only the power-law is adopted. 
}
\label{fig:lf}
\end{figure*}

The three observed LFs are shown in Figure~\ref{fig:lf}, which consist of 302 (Fornax-field), 85 (NGC\,1399-field) and 82 (Fornax-GC) sources, respectively. 
We fit the LFs using a canonical power-law model, $dN/dL \propto L^{-\alpha}$. 
This model is corrected for incompleteness and Eddington bias, following the procedure of \citet{Wang2004ApJ} and \citet{Zhu2018} and taking into account the source spatial distribution. To zeroth order, a flat distribution is assumed for Fornax-field and Fornax-GC, and a S{\'e}rsic profile is assumed for NGC\,1399-field (Section~\ref{sec:spatial}).
For Fornax-field, we also include a fixed component, according to the $logN-logS$ relation of \citet{Georgakakis2008MNRAS}, to account for the CXB contribution (dotted curve in Figure~\ref{fig:lf}). 
The Fornax-field LF shows a clear excess above the CXB level at photon fluxes $\lesssim2\times10^{-6}{\rm~ph~cm^{-2}~s^{-1}}$. 
The best-fit power-law slope is obtained by minimizing the $C$-statistic \citep{Cash1979}. 
As summarized in Table~\ref{tab:LF}, the fit result indicates a rather steep LF of the Fornax-field ($\alpha = 2.54^{+0.38}_{-0.29}$, 90\% confidence level), which is statistically consistent with that of the NGC\,1399-field ($\alpha = 2.35^{+0.29}_{-0.22}$).
On the other hand, the Fornax-GC LF is significantly flatter, with $\alpha = 1.94^{+0.15}_{-0.14}$.
That the GC-LMXBs have a flatter LF than their field counterparts is consistent with previous work (e.g., \citealp{Kim2009,Zhang2011A&A}).


We note that in the above analysis we have ignored $\sim$100 sources detected only in the $S$- or $H$-band, chiefly to minimize the uncertainty in the CXB contribution and the LF. Some of these sources can also belong to the intra-cluster populations. 

\section{Discussion} \label{sec:discussion}
We have investigated the spatial and flux distributions of X-ray point sources in Fornax, out to a projected distance of $\sim 30\arcmin$ ($\sim$ 180 kpc) from the cluster center. 
After accounting for sources of well-established origins, i.e., those associated with the member galaxies (in particular the BCG), known GCs and the CXB, significant excess is found over a substantial radial range (Figure~\ref{fig:rp}).
Below we discuss the possible nature of these excess sources, in close comparison with their counterparts previously identified in the Virgo cluster \citep{Hou2017ApJ}.


The first and obvious possibility is that some of the excess sources are still GC-LMXBs, despite our comprehensive effort in identifying them.
Indeed, in the case of Virgo, among the $\sim$120 excess sources with $L_{\rm X} \gtrsim 10^{38}{\rm~erg~s^{-1}}$, $\sim$30\% can be attributed to GC-LMXBs\footnote{In \citet{Hou2017ApJ}, the excess was identified without excluding GC-LMXBs known {\it a pirior}. The current choice of precluding GC-LMXBs to our best knowledge therefore means a more stringent definition of excess sources.} (\citealp{Hou2017ApJ}). 
In Fornax, we have found GC-LMXBs out to a radius of 25\farcm6, and GCs have been detected even beyond the edge of our FoV \citep{Bassino2006A&A,Cantiello2018A&A}. 
As addressed in Section~\ref{subsec:counterpart}, the majority of the GC-LMXBs are found in bright, massive GCs, and only $\lesssim$20 X-rays sources associated with faint, undetected GCs are likely to have escaped our identification.
Moreover, the LF of Fornax-GC sources is much flatter than that of the Fornax-field sources (Section~\ref{sec:lf}), suggesting that the two sets of sources have a different origin.  Hence we conclude that any unidentified GC-LMXBs would have only a minor contribution to the excess sources. 

The second plausible scenario to consider is supernova-kicked LMXBs, favorably consisting of an accreting NS.
If the NS received a kick velocity greater than the host galaxy's escape velocity (on the order of 100~km~s$^{-1}$), which is due to its parent supernova explosion, and the binary system in which the NS resides survived this kick, an NS-LMXB might be later found at the galaxy outskirt or the intra-cluster space \citep{Brandt1995,Zuo2008}.
While such a scenario could be generic, supernova-kicked LMXBs were considered to have a negligible contribution to the excess sources found in Virgo \citep{Hou2017ApJ}, because the latter were detected with $L_{\rm X} \gtrsim 2\times10^{38}{\rm~erg~s^{-1}}$, i.e., above the Eddington limit for NS-LMXBs.  
Given the current detection limit of $\sim$$3\times10^{37}{\rm~erg~s^{-1}}$ (Figure~\ref{fig:hr}), supernova-kicked LMXBs are likely present in the excess sources. 
\citet{Zhang2013} proposed that about half of the excess sources (with $L_{\rm X} \gtrsim 10^{37}{\rm~erg~s^{-1}}$) found in their sample of ETGs can be attributed to supernova-kicked LMXBs, which on average have an abundance of $\sim$0.5 source per $10^{10}{\rm~M_{\odot}}$ (with $L_{\rm X} \geq 5\times10^{37}{\rm~erg~s^{-1}}$, normalized to the host galaxy's stellar mass). 
Assuming a similar scaling holds in Fornax, we estimate that $\lesssim$34 supernova-kicked LMXBs are amongst our excess sources, a value obtained by summing up the 29 member galaxies including NGC\,1399.

The third possibility is X-ray binaries associated with a more diffuse stellar population, either the extended stellar halo of NGC\,1399 \citep{Iodice2016ApJ} or the ICL \citep{Iodice2017ApJ}. 
The BCG halo grows primarily from mergers of infalling galaxies, whereas the ICL is built up mainly through stars tidally stripped from the outskirts of large galaxies or tidally disrupted low-mass galaxies orbiting in the cluster potential. 
Both being the result of hierarchical clustering, these two components are physically connected to each other, and it is noteworthy that a clear cut between them is rather difficult with morphological information only. 
Nevertheless, consisting of predominantly old stellar populations, both the BCG halo and ICL are expected to host LMXBs. 
In Section~\ref{sec:spatial}, we have attempted a spatial decomposition of the BCG halo contribution by fitting the observed surface density profile with a presumed exponential form, finding that this halo does appear to be the dominant component over the range of $R \approx 3-8 R_{\rm e}$ (Figure~\ref{fig:rp}), potentially accounting for $\sim$110 excess sources. 
In a recent study of X-ray sources in M87, the central giant elliptical galaxy of Virgo, \citet{Luan2018} found that the X-ray source surface density profile is consistent with the stellar halo distribution out to a projected radius of $\sim$100 kpc. 
\citet{vanHaaften2018} also reported an excess of X-ray sources in the halo of NGC\,4472, a group-central giant elliptical galaxy, although these authors essentially referred to the excess as sources without a known optical counterpart. 

The exponential halo, however, cannot solely account for the detected excess.
Subtracting the best-fitted halo contribution still leaves a residual of $\sim$ 74 sources (or 40\% of all excess sources) spreading over a large radial range (Figure~\ref{fig:rp}).
This residual is likely related to the putative ICL.
Based on the FDS, \citet{Iodice2017ApJ} found clear evidence for the presence of ICL in the core of Fornax, 
the bulk of which lies between NGC\,1387, NGC\,1379 and NGC\,1381 (a region $\sim$10\arcmin--40\arcmin~west of NGC\,1399) and possibly arises from tidal disruption of these three galaxies.
Using the ICL $g$-band luminosity of $8.3\times10^{9}{\rm~L_\odot}$ measured by \citet{Iodice2017ApJ}, and adopting the values of $M/L$ and $\epsilon_{\rm X}$ obtained in Section~\ref{sec:spatial}, we estimate that $\sim$20 LMXBs could be associated with the ICL in this particular region. 
As a related remark, some GCs (and any associated GC-LMXBs) might also be classified as part of the ICL. Indeed, \citet{Iodice2017ApJ} found the aforementioned ICL to be spatially coincident with a previously known overdensity of blue GCs \citep{D'Abrusco2016ApJ}.

The last possible origin of the excess we consider, which is also closely related to hierarchical evolution in cluster environment, is the relic of tidally stripped nucleated galaxies \citep{Ferguson1994}.
In particular, relic galactic nuclei can manifest themselves as UCDs (e.g., \citealp{Liu2015}),   
whose typical luminosities and sizes (hence stellar densities) are intermediate between the classical GCs and dwarf elliptical galaxies \citep{Bruns2012}. 
Empirically $\sim$3\% of the known UCDs have an X-ray counterpart with $L_{\rm X} \gtrsim 10^{37}{\rm~erg~s^{-1}}$ \citep{Hou2016}, which are most likely LMXBs formed via stellar encounters, as in the case of GCs.
Alternatively, a massive BH, if existed at the center of the UCD, can also give rise to detectable X-rays. Such a case might be relevant to M60-UCD1 \citep{Seth2014} and M59-UCD3 \citep{Ahn2018ApJ}, two very massive UCDs both showing strong evidence for a central massive BH.   
Among the 144 UCDs currently known in Fornax, we have identified fifteen with an X-ray counterpart, but due to the somewhat ambiguous classification of UCDs, 13 of these have also been considered a GC (Section~\ref{subsec:counterpart}). 
In any case, UCDs and relic galactic nuclei are unlikely to have a substantial contribution to the excess sources.
In passing, we note that no X-ray counterpart (with a 3\,$\sigma$ upper limit of $L_{\rm X} < 2\times10^{37}{\rm~erg~s^{-1}}$) is found for Fornax UCD3, which is recently suggested to host a central BH of 3.5-million solar masses \citep{Afanasiev2018}.
A detailed study of the X-ray properties of the UCDs, in parallel with the identified GC-LMXBs in Fornax, will be presented elsewhere.  


\section{Concluding Remarks} \label{sec:summary}
Using extensive {\it Chandra} observations, we have studied the X-ray sources in the inner $\sim$30\arcmin~($\sim$180 kpc) of the Fornax cluster. Our main results include:
\begin{itemize}
\item{We detect a total of 1279 independent X-ray sources, among which 1177 are detected in the 0.5--8 keV band down to a limiting luminosity of $\sim$$3\times10^{37}{\rm~erg~s^{-1}}$. By cross-correlating various optical catalogs, we identify 18 foreground sources, 270 GCs, 15 UCDs and 4 nuclear sources. A source catalog is presented for future reference.}
\item{From the radial surface density profile, we statistically identify $\sim$183 excess sources beyond three times the effective radius of NGC\,1399, with respect to the expected CXB.}
\item{The luminosity function of the excess sources is significantly steeper than that of the GC-LMXBs, suggesting that any unidentified GCs would have only a minor contribution to the excess sources.}
\item{We find that LMXBs associated with either the extended stellar halo of NGC\,1399 or the ICL can be responsible for the majority of the excess sources. Supernova-kicked LMXBs may account for a small but non-negligible fraction of the excess.}
\end{itemize}

The above findings provide strong evidence for the presence of intra-cluster X-ray sources in Fornax, the second unambiguous case for a galaxy cluster after Virgo. 
Owing to the low surface brightness, it is a persistent challenge to detect ICL with conventional optical imaging. Discrete sources, such as GCs, planetary nebulae \citep{Theuns1997} and novae \citep{Neill2005}, have been proposed as a complementary tracer for the ICL. 
The detection of intra-cluster X-ray sources in Virgo \citep{Hou2017ApJ} and Fornax (this work) now opens up a new window to study the ICL. 
The rather steep power-law slope of the excess sources in Fornax indicates that there could be many unresolved, fainter sources below the current detection limit, which can be probed with deeper {\it Chandra} observations.
In the case of Virgo, where the large sky area renders it impractical to conduct a full mapping with {\it Chandra} or {\it XMM-Newton}, 
the upcoming all-sky survey by the eROSITA mission \citep{Merloni2012} is expected to find at least 10 times more intra-cluster X-ray sources with $L_{\rm X} \gtrsim 2\times10^{38}{\rm~erg~s^{-1}}$.
Such sources may form a statistically meaningful sample for exploring the physical properties and formation history of the ICL in the two clusters.

\acknowledgements
This work is supported by National Key Research and Development Program of China (2017YFA0402703) and National Science Foundation of China under grants 11473010 and J1210039. X.J. acknowledges support from Top-notch Academic Programs Project of Jiangsu Higher Education Institutions.
We thank Yisi Yang for his help at the early stage of data analysis. 


\clearpage
\begin{deluxetable}{cccccc}
\tabletypesize{\footnotesize}
\tablewidth{0pt}
\tablecaption{Log of {\it Chandra} observations}
\tablehead{
\colhead{ObsID} &
\colhead{Instrument} &
\colhead{RA} & 
\colhead{DEC} &
\colhead{Exposure} &
\colhead{Start date} \\
\colhead{(1)} &
\colhead{(2)} &
\colhead{(3)} &
\colhead{(4)} &
\colhead{(5)} &
\colhead{(6)}
}
\startdata
624 & ACIS-S & 03 39 34.70 & -35 25 50.00 & 43.6 & 1999-12-15 \\
319 & ACIS-S & 03 38 29.40 & -35 27 00.40 & 56.0 & 2000-01-18 \\
239 & ACIS-I & 03 38 29.40 & -35 27 00.40 & 3.6 & 2000-01-19 \\
2942 & ACIS-S & 03 38 52.00 & -35 35 34.00 & 29.2 & 2003-02-13 \\
4168 & ACIS-I & 03 36 59.85 & -35 29 39.84 & 45.6 & 2003-05-20 \\
4169	& ACIS-I & 03 37 09.50 & -35 44 00.06 & 37.4 & 2003-05-21 \\
4171 & ACIS-I & 03 38 11.20 & -35 41 54.20 & 45.0 & 2003-05-23 \\
4170	 & ACIS-I & 03 37 14.40 & -35 13 27.84 & 40.6 & 2003-05-24 \\
4172	 & ACIS-I & 03 38 25.56 & -35 25 42.60 & 44.5	& 2003-05-26  \\
4173 & ACIS-I & 03 38 37.01 & -35 09 27.72 & 45.1 & 2003-05-26 \\
4174	 & ACIS-I & 03 38 49.58 & -35 34 36.34 & 45.7 & 2003-05-28 \\
4175	 & ACIS-I & 03 39 30.51 & -35 45 22.03 & 54.6 & 2003-05-29 \\
4176 & ACIS-I & 03 39 44.60 & -35 29 08.92 & 46.0 & 2003-05-31 \\
4177	 & ACIS-I & 03 39 55.80 & -35 12 56.16 & 38.9 & 2003-06-01 \\ 
3949 & ACIS-S & 03 40 10.50 & -35 37 38.00 & 54.6 & 2003-10-21 \\
9798	 & ACIS-S & 03 38 51.00 & -35 34 31.00 & 18.3 & 2007-12-24 \\
9799 & ACIS-S & 03 38 51.00 & -35 34 31.00 & 21.3 & 2007-12-27 \\
9530 & ACIS-S & 03 38 29.00 & -35 27 01.40 & 59.4 & 2008-06-08 \\
14527 & ACIS-S & 03 38 29.10 & -35 27 03.00 & 27.8 & 2013-07-01 \\
16639 & ACIS-S & 03 38 29.10 & -35 27 03.00 & 29.7 & 2014-10-12 \\
16231 & ACIS-S & 03 39 01.20 & -35 35 18.60 & 60.5 & 2014-10-20 \\
17541 & ACIS-S & 03 39 01.20 & -35 35 18.60 & 24.7 & 2014-10-23 \\
16234 & ACIS-S & 03 39 22.32 & -35 38 42.00 & 90.9 & 2014-10-30 \\
17540 & ACIS-S & 03 39 01.20 & -35 35 18.60 & 28.5 & 2014-11-02 \\
16233 & ACIS-S & 03 39 04.08 & -35 35 31.20 & 98.8 & 2014-11-09 \\
17548 & ACIS-S & 03 39 04.08 & -35 35 31.20 & 48.2 & 2014-11-11 \\
16232 & ACIS-S & 03 39 04.08 & -35 35 31.20 & 69.1 & 2014-11-12 \\
17549 & ACIS-S & 03 38 51.26 & -35 38 19.28 & 61.7 & 2015-03-28 \\
14529 & ACIS-S & 03 38 29.10 & -35 27 03.00 & 31.6 & 2015-11-06 \\
\enddata
\tablecomments{(1) Observation ID; (2) Instrument; (3)-(4) Epoch 2000 Coordinates of the aim-point; (5) Cleaned exposure, in the units of ks; (6) Start date of the observation, in the form of yyyy-mm-dd.}
\label{tab:journal}
\end{deluxetable}

\begin{deluxetable}{cccccccccc}
\tabletypesize{\footnotesize}
\tablecaption{Catalog of Detected X-ray Sources in the Fornax Cluster}
\tablewidth{0pt}
\tablehead{
\colhead{No.} &
\colhead{R.A.} & 
\colhead{Dec} &
\colhead{Pos. err} &
\colhead{$S_{0.5-2}$}ß &
\colhead{$S_{2-8}$} &
\colhead{$S_{0.5-8}$} &
\colhead{$F_{0.5-8}$} &
\colhead{HR} &
\colhead{Note} \\
\colhead{(1)} &
\colhead{(2)} &
\colhead{(3)} &
\colhead{(4)} &
\colhead{(5)} &
\colhead{(6)} &
\colhead{(7)} &
\colhead{(8)} &
\colhead{(9)} &
\colhead{(10)}
}
\startdata
1 & 54.07722 & -35.48106 & 1.60 & $<5.9$ & $<22.0$ & $14.8^{+4.5}_{-5.9}$ & $53.9^{+16.4}_{-21.6}$ & $0.58^{+0.42}_{-0.12}$ & ... \\
2 & 54.08801 & -35.55983 & 0.42 & $216.8^{+14.4}_{-17.2}$ & $54.3^{+8.0}_{-8.0}$ & $269.3^{+17.2}_{-18.1}$ & $980.3^{+62.8}_{-65.9}$ & $-0.59^{+0.05}_{-0.06}$ & ... \\
3 & 54.11255 & -35.51206 & 0.75 & $14.3^{+3.6}_{-4.1}$ & $<16.2$ & $23.0^{+4.6}_{-5.3}$ & $83.6^{+16.6}_{-19.1}$ & $-0.25^{+0.25}_{-0.20}$ & ... \\
4 & 54.12001 & -35.37550 & 1.06 & $35.4^{+5.9}_{-7.0}$ & $<31.4$ & $54.1^{+7.2}_{-8.9}$ & $197.0^{+26.1}_{-32.4}$ & $-0.29^{+0.15}_{-0.15}$ & ... \\
5 & 54.12192 & -35.54933 & 0.60 & $39.2^{+5.5}_{-7.2}$ & $<14.4$ & $46.3^{+5.3}_{-8.7}$ & $168.7^{+19.5}_{-31.6}$ & $-0.69^{+0.10}_{-0.14}$ & ... \\
6 & 54.13016 & -35.44953 & 0.79 & $10.7^{+2.6}_{-4.4}$ & $<15.4$ & $18.8^{+4.0}_{-5.6}$ & $68.4^{+14.5}_{-20.3}$ & $-0.13^{+0.25}_{-0.28}$ & G \\
7 & 54.13259 & -35.29532 & 1.05 & $<167.6$ & $<42.7$ & $143.1^{+18.0}_{-18.7}$ & $521.0^{+65.7}_{-68.2}$ & $-0.64^{+0.08}_{-0.12}$ & N \\
8 & 54.13600 & -35.29274 & 1.08 & $<114.8$ & $<59.0$ & $114.7^{+13.3}_{-18.7}$ & $417.7^{+48.5}_{-68.0}$ & $-0.33^{+0.12}_{-0.14}$ & G \\
9 & 54.13765 & -35.59933 & 0.85 & $43.2^{+6.0}_{-5.7}$ & $<25.7$ & $56.8^{+6.5}_{-6.8}$ & $206.8^{+23.5}_{-24.9}$ & $-0.45^{+0.10}_{-0.12}$ & ... \\
10 & 54.14037 & -35.41325 & 0.56 & $35.1^{+5.8}_{-6.5}$ & $22.3^{+3.5}_{-6.1}$ & $57.2^{+7.0}_{-8.3}$ & $208.4^{+25.4}_{-30.3}$ & $-0.21^{+0.11}_{-0.16}$ & ... \\
\enddata
\tablecomments{(1) Source ID, in order of increasing R.A.; (2)-(3) Right Ascension and Declination (J2000) of source centroid; (4) Position uncertainty in arc-seconds; (5)-(7) The 0.5-2~keV, 2-8~keV and 0.5-8~keV photon flux, in units of $10^{-7}\;\!{\rm ph\;\!cm^{-2}\;\!s^{-1}}$;
(8) The 0.5-8~keV energy flux, in units of $10^{-16}\;\!{\rm erg\;\!cm^{-2}\;\!s^{-1}}$, converted from 0.5-8~keV photon flux by assuming an absorbed power-law spectrum with photon-index of 1.7 and column density $10^{21}\;\!{\rm cm^{-2}}$;
(9) Hardness ratio between the 0.5--2 keV and 2--8 keV bands;
(10) ``F" denotes foreground stars, ``G" denotes sources spatially coincident with known GCs, ``N" denotes nuclear source of a member galaxy, and ``U" denotes sources spatially coincident with known UCDs.
Quoted errors are at 1\,$\sigma$ confidence level, while 3\,$\sigma$ upper limits are given in the case of non-detection in a given band.
(Only a portion of the full table is shown here to illustrate its form and content.)
}
\label{tab:sourcecatalog}
\end{deluxetable}

\begin{deluxetable}{cccccc}
\tablewidth{0pt}
\tablecaption{Fornax Cluster Galaxies within the {\it Chandra} FoV}
\tabletypesize{\footnotesize}
\tablehead{
\colhead{Galaxy Name} &
\colhead{Other Name} & 
\colhead{R. A.} &
\colhead{Dec} &
\colhead{$R_{\rm e}$} &
\colhead{Source ID}\\
\colhead{(1)} &
\colhead{(2)} &
\colhead{(3)} &
\colhead{(4)} &
\colhead{(5)} &
\colhead{(6)} 
}
\startdata
FCC\, 170&NGC1381&54.1317&-35.2953& 12.9 & 7 \\
FCC\, 171& &54.1557&-35.3846& 18.0 & \\
FCC\, 175& &54.1779&-35.4341&  7.7 & \\
FCC\, 182& &54.2261&-35.3729& 11.4 & \\
FCC\, 184&NGC1387 &54.2369&-35.5066& 50.1 & 63\\
FCC\, 188& &54.2689&-35.5886& 12.1 & \\
FCC\, 190&NGC1380B&54.287&-35.1937& 16.3 & \\
FCC\, 191&  &54.2913&-35.3854&  6.3\\
FCC\, 193&NGC1389&54.2987&-35.7443& 20.1 & \\
FCC\, 194& &54.3243&-35.6972&  6.4 & \\
FCC\, 196& &54.3909&-35.8277& 10.2 & \\
FCC\, 197& &54.4203&-35.2948&  6.3 & \\
FCC\, 202&NGC1396&54.5267&-35.4383&  9.8 & \\
FCC\, 207& &54.5797&-35.1274&  8.5 & \\
FCC\, 208& &54.5778&-35.5290& 11.7 & \\
FCC\, 211& &54.5889&-35.2582&  5.6 & \\
FCC\, 213&NGC1399&54.6215&-35.4506& 49.1 & 527 \\
FCC\, 214& &54.652&-35.8324&  6.2 & \\
FCC\, 215& &54.6562&-35.7555&  7.4 & \\
FCC\, 218& &54.6887&-35.2645&  7.6 & \\
FCC\, 219&NGC1404&54.7171&-35.5938& 20.0 & 841 \\
FCC\, 220& &54.7292&-35.2352&  5.6 & \\
FCC\, 222& &54.8053&-35.3697& 14.5 & \\
FCC\, 223& &54.8311&-35.7234& 16.6 & \\
FCC\, 227&  &54.9587&-35.5211&  7.3 & \\
FCC\, 228& &54.9636&-35.3206&  9.2 & \\
FCC\, 229& &54.9799&-35.6609&  6.8 & \\
FCC\, 235&NGC1427A&55.0383&-35.6231& 36.3 & \\
FCC\, 241& &55.097&-35.2744& 15.9 & \\
\enddata
\tablecomments{(1) Member galaxies in the Fornax Cluster Catalog \citep{Ferguson1989AJ}; (2) Other names of the galaxy; 
(3)-(4) J2000 coordinates of the galactic center; (5) Effective radius in arc-seconds. The value for NGC\,1399 is adopted from \citet{Iodice2016ApJ}; (6) ID of the identified nuclear X-ray source, as listed in Table \ref{tab:sourcecatalog}. }
\label{tab:FCC}
\end{deluxetable}

\begin{deluxetable}{lccc}
\tabletypesize{\small}
\tablewidth{-30pt}
\tablecaption{Fitted Luminosity Functions}
\tablehead{
\colhead{Source}&
\colhead{$\alpha$}&
\colhead{$N(> L_{\rm ref})$} &
\colhead{$C$/d.o.f.} 
}
\startdata
Fornax-field & $2.54^{+0.38}_{-0.29}$ &  306 & 41.7/23\\
Fornax-GC & $1.94^{+0.15}_{-0.14} $ & 199 & 5.3/7  \\
NGC\,1399-field & $2.35^{+0.29}_{-0.22}$ & 126  &11.8/6 \\
\enddata
\tablecomments{
The luminosity functions are fitted by a power-law: $dN/dL \propto L^{-\alpha}$, where the normalization is expressed as the number of sources more luminous than $L_{\rm ref} = 5\times10^{37}{\rm~erg~s^{-1}}$. Quoted errors for the slope $\alpha$ are at 90\% confidence level.
In the case of Fornax-field sources, an additional, fixed component for the CXB has been included in the fit.
}
\label{tab:LF}
\end{deluxetable}

\end{document}